\documentclass[twocolumn,prl,showpacs,superscriptaddress]{revtex4}

\usepackage{graphicx}
\usepackage{bm}

\begin{document}

\title{Orbital mixing and nesting in
the bilayer manganites La$_{2-2x}$Sr$_{1+2x}$Mn$_2$O$_7$}

\author{R. Saniz}
\altaffiliation[Present address: ]{Departement Fysica,
Universiteit Ant\-wer\-pen, Groenenborgerlaan 171, B-2020 Antwerpen,
Belgium.}
\affiliation{Department of Physics and Astronomy,
Northwestern University, Evanston, Illinois  60208}

\author{M. R. Norman}
\affiliation{Materials Science Division, Argonne National Laboratory,
Argonne, Illinois  60439}

\author{A. J. Freeman}
\affiliation{Department of Physics and Astronomy,
Northwestern University, Evanston, Illinois  60208}

\date{\today}


\begin{abstract}

A first principles study of
La$_{2-2x}$Sr$_{1+2x}$Mn$_2$O$_7$ compounds for doping levels
$0.3\leq x\leq 0.5$
shows that the low energy electronic structure of the majority spin carriers 
is determined by strong momentum dependent interactions
between the Mn $e_g$  $d_{x^2-y^2}$ and $d_{3z^2-r^2}$ orbitals, which
in addition to an $x$ dependent Jahn-Teller distortion,
differ in the ferromagnetic and antiferromagnetic phases.
The Fermi surface exhibits nesting behavior that is
reflected by peaks in the static susceptibility,
whose positions as a function of momentum
have a non-trivial dependence on $x$.

\end{abstract}

\maketitle

One of the most studied series of compounds over the past decade,
having being scrutinized using a wide range of experimental
methods~\cite{kimura00},
is that of the bilayer manganites La$_{2-2x}$Sr$_{1+2x}$Mn$_2$O$_7$,
with new findings of their basic
properties continuing to emerge~\cite{mannella07,dejong07}.
Their interest is due to several
reasons. It is well known that
they exhibit a colossal magnetoresistive effect around critical
temperatures of 100-200 K, depending on the doping level
$x$~\cite{kimura00}.  A proper
understanding of this phenomenon should point the way to tailoring materials
with higher critical temperatures, which could potentially lead to magnetic
devices that would outperform
present ones based on the giant magnetoresistive effect.
On the fundamental side, it is thought that the properties of these
materials are due to an interplay of structural, orbital, and magnetic
degrees of freedom, a trait shared by several strongly correlated materials,
including the superconducting cuprates~\cite{dagotto05}.
In the case of the manganites, there is no consensus as yet on exactly how
these different degrees of freedom lead to the observed properties.
Elucidating this will no doubt lead to an advance in our understanding
of related phenomena in the broader context of
higher complexity condensed matter systems.

One of the reasons for the current state of affairs is that the ground state
properties of the bilayer manganites are not completely understood, even
at low temperatures. From the theoretical standpoint, there are few
published reports of {\it ab initio} calculations of the electronic
structure of these compounds~\cite{huang00}.
While these studies have provided an important framework for the
analysis of experiments, several observations remain to be
addressed. For instance, it is generally recognized that
it is the character of the occupied Mn $e_g$ states that determines
the key properties of these systems, such as the conductivity and the nature
of the magnetic order.
Thus, important efforts have been made to understand the
orbital polarization of these
states~\cite{koizumi01,li04,barbiellini05,koizumi06}, but the problem
has only been partially addressed from an {\it ab initio}
perspective~\cite{mijnarends07}.
In this work, we present a first principles study of the conducting states
of the bilayer manganites for hole doping levels $0.3\leq x\leq 0.5$.
We show that the $d_{x^2-y^2}$ and
$d_{3z^2-r^2}$ orbitals strongly mix, with the mixing
depending not only on $x$, but also on the direction
in {\bf k}-space.
This has a pronounced impact on the
Fermi surface topology and nesting, long
suggested to play an important role in these
materials~\cite{chuang01,sun06,kim07}. This is illustrated by
calculating the static susceptibility $\chi({\bf q})$, which
indeed shows peaks due to nesting.
The implications of our work in relation to recent experiments are
discussed.

We employ the highly precise all-electron full-potential linearized
augmented plane wave (FLAPW) implementation of density functional
theory~\cite{wimmer81}, with the generalized gradient
approximation (GGA)~\cite{perdew97,comment2}
for the exchange-correlation potential. For $x<0.5$, we use the ``charged
cell" approximation, i.e., increasing the valence electron number by the
required amount and adding a uniform positive charge for
neutrality~\cite{comment}. The calculations were carried out
for the observed magnetic phases:  ferromagnetic
(FM) for $x\leq 0.4$ and antiferromagnetic (AF) for $x\geq 0.45$. In our
calculations, the AF
phase is given by ferromagnetic MnO$_2$ planes that are coupled
antiferromagnetically within a bilayer as well as
between adjacent bilayers.  Though in experiment
the latter coupling is ferromagnetic in the A-type AF order, it is known to
be very weak~\cite{kimura00}, and so should not influence our
results~\cite{comment2}.
As for technical aspects of the calculations, convergence was assured with
respect to muffin-tin radii, {\bf k}-point mesh [1164 points in the
irreducible wedge of the Brillouin zone], and energy cut-offs. The
structural parameters for each $x$ ($I4/mmm$) are from
Ref.~\onlinecite{kubota00} at 10 K.  We assume 
La is in the $2b$ and Sr in the $4e$ sites~\cite{seshadri97}.

\begin{figure}
\includegraphics[width=\hsize]{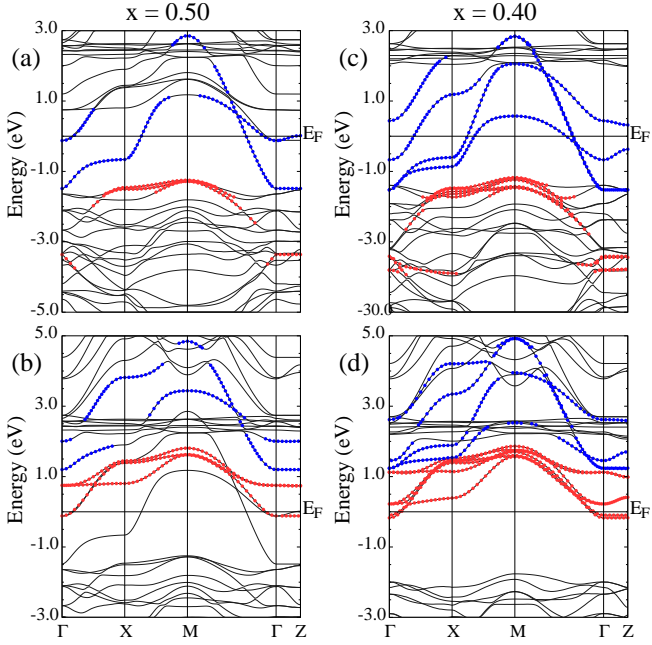}
\caption{\label{fig1} (color online) 
Band structure of LaSr$_2$Mn$_2$O$_7$ ($x=0.5$, AF phase) for
$\uparrow$ [$\downarrow$] spin in panel (a) [(b)], and of 
La$_{1.2}$Sr$_{1.8}$Mn$_2$O$_7$ ($x=0.4$, FM phase) for $\uparrow$
[$\downarrow$] spin in panel
(c) [(d)]. The gray (red) and black (blue) dots highlight states of
dominant Mn $t_{2g}$ and
$e_g$ character, respectively.}
\end{figure}

As indicated above, key to the understanding of the 
bilayer manganites is the relative role of the $e_g$ states
($d_{x^2-y^2}$ and $d_{3z^2-r^2}$)~\cite{kimura00}. To illustrate
these states, we refer to the band structure plots
in Fig.~\ref{fig1} (simple tetragonal symmetry notation is used).
Two cases are shown, namely $x=0.5$ (AF) and $x=0.4$ (FM).
The states with more
than 50\% Mn d
character are highlighted with dots.
In the AF case,
there is Kramers degeneracy, so the two spin states are interchanged for the
second Mn atom
type (there is only one Mn atom type in the FM case). In the
following, we focus on the $e_g$ states.

The basic electronic structure
near the Fermi energy ($E_{\rm F}$) can be
understood as arising from the two Mn $d_{x^2-y^2}$ and two Mn $d_{3z^2-r^2}$
orbitals per
bilayer.  Ignoring orbital mixing and hybridization with other states, in the
FM case the energy of each orbital would
be of the form $\epsilon_i \pm X \pm \Delta$, where $\epsilon_i$ is the
unpolarized atomic orbital energy ($i=1,2$),
2$X$ the exchange splitting, and 2$\Delta$ the bilayer splitting, resulting
in eight states altogether.  In contrast, in the AF
case the energy of each orbital would be of the form
$\epsilon_i \pm \sqrt{X^2 + \Delta^2}$,
resulting in four Kramers degenerate states.  Since $X \gg \Delta$, the
bilayer splitting is essentially quenched in the AF case.  Consequently,
as we show below,
there is only one barrel centered at M in the AF case, as opposed to two in
the FM case.
Moreover, for the AF case, unlike the FM one, we do not expect a strong
intensity
modulation of the photoemission signal as a function of photon energy due to
bilayer splitting.

\begin{figure}
\includegraphics[width=\hsize]{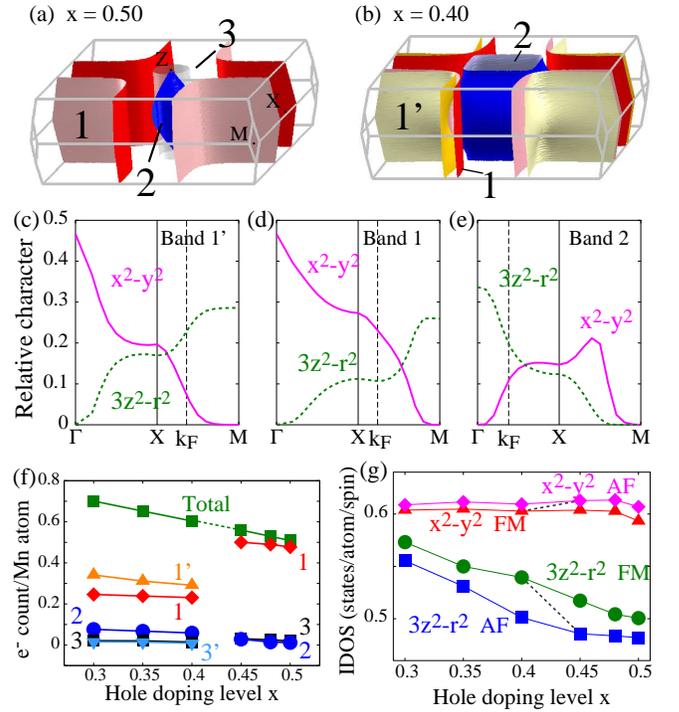}
\caption{\label{fig2} (color online)
Fermi surface for (a)
$x=0.5$ (AF)  and (b) $x=0.4$ (FM). 
(c), (d), and (e): Orbital character of the states
along the $\Gamma$-X-M line for the $e_g$ bands crossing $E_{\rm F}$
for $x=0.4$. (f) Luttinger count of the different
Fermi surface sheets, and (g) integrated density of states (IDOS)
for $d_{x^2-y^2}$ and $d_{3z^2-r^2}$ as a function of doping,
imposing either AF or FM order. Note the crossover in orbital
population due to the change in magnetic order (dotted lines).}
\end{figure}

In the actual calculations the two $e_g$ orbitals are mixed.
This is directly reflected in the Fermi surface topology.
In Fig.~\ref{fig2}(a), we show the Fermi surface for $x=0.5$ (AF),
and in Fig.~\ref{fig2}(b) for
$x=0.4$ (FM,  majority spin). In the AF case, there is a
hole-like barrel centered at M of mainly $d_{x^2-y^2}$ symmetry (labeled 1),
a prolate electron-like sheet around $\Gamma$ of $d_{3z^2-r^2}$ symmetry
(labeled 2), and
a (minority spin) cylindrical sheet centered at $\Gamma$ of $d_{xy}$ symmetry
(labeled 3).
In the FM phase, there is a hole-like barrel of dominant $d_{x^2-y^2}$
symmetry centered at M
(labeled 1), and two barrels of mixed $d_{x^2-y^2}$
and $d_{3z^2-r^2}$ symmetry: one hole-like, centered at M 
(labeled 1'), and the other
electron-like, centered at $\Gamma$ (labeled 2)~\cite{comment0}.

The mixing is shown by an analysis
of the character of the states for bands
crossing $E_{\rm F}$. For example,
in Fig.~\ref{fig2}(c) we show the character of the lower $e_g$ band (1')
for $x=0.4$ (FM) and {\bf k} points along the $\Gamma$-X-M
directions. It starts with
$d_{x^2-y^2}$ character at $\Gamma$, but when reaching
X, the mixing with $d_{3z^2-r^2}$ is about 50\%.
At $k_{\rm F}$, roughly halfway between X and M, the character becomes
dominantly $d_{3z^2-r^2}$. The middle band (1), in contrast
[Fig.~\ref{fig2}(d)],
has a fairly dominant $d_{x^2-y^2}$ character up to $k_{\rm F}$, at
about a quarter of the X-M distance. Band 2, in turn [Fig.~\ref{fig2}(e)],
starts out
at $\Gamma$ as $d_{3z^2-r^2}$ but is strongly mixed at $k_{\rm F}$.
Thus, it is inaccurate to characterize the barrels around the
M point as `$x^2-y^2$', and the electron pocket around $\Gamma$ as
`$3z^2-r^2$'.
Our results for the other doping levels in
the FM phase show that the orbital mixing increases with decreasing
$x$, and that the bilayer splitting becomes stronger.  
A similar analysis shows that in the AF case, 
the lower $e_g$ band (1) is of dominant
$d_{x^2-y^2}$ character, but with strong mixing at  $k_{\rm F}$ along
X-M. The upper band (2) is $d_{3z^2-r^2}$ like.

\begin{figure}
\includegraphics[width=0.49\hsize]{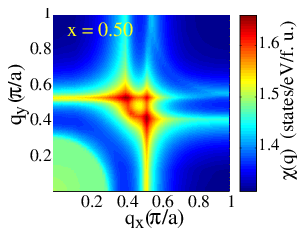}
\includegraphics[width=0.49\hsize]{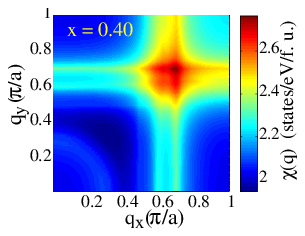}
\caption{\label{fig3} (color online)
The static susceptibility, $\chi({\bf q})$, for
${\bf q}=(q_x,q_y,2\pi/c)$, for $x=0.5$ (AF) and $x=0.4$ (FM).}
\end{figure}

Of further relevance to experiment is the occupation level of the
bands crossing $E_{\rm F}$. Thus, we calculated the carrier number by
integrating the volume enclosed by the different Fermi surfaces
(the Luttinger count). We used a fine mesh of 32340 {\bf k}-points in the
irreducible wedge, and the results as a function of $x$ are shown in
Fig.~\ref{fig2}(f).
In the AF phase, sheet 1 is obviously dominant.
In the FM
phase, the Kramers degeneracy is lifted,
giving rise to sheets 1 and 1'. The increased orbital
mixing lowers the energy, particularly in the case of states 1', allowing
the charge to increase as $x$ decreases.
In the FM phase there is only one sheet 2 because the antibonding counterpart
is empty (due to the large bilayer splitting of $3z^2-r^2$ in the FM phase).
Thus, the electron pocket around $\Gamma$ roughly doubles
in size compared to the AF phase.
Note that for $0.4<x\alt0.45$ the
actual phase is a canted antiferromagnet~\cite{kubota00}, so the
bilayer splitting is probably reduced gradually with $x$.
But clearly, AF coupling acts to quench the bilayer splitting,
with dramatic effects, particularly for the $d_{3z^2-r^2}$ states.
This is in line with very recent photoemission data that indicate
a collapse of bilayer splitting near $x$=0.4~\cite{jozwiak08,comment1}.

In Fig.~\ref{fig2}(g)
we plot the $e_g$ orbital occupation counts (estimated by integrating
the muffin-tin projected density of states) as a function of $x$,
imposing either a FM or AF phase.
The $d_{x^2-y^2}$ orbital population is not strongly affected across the
FM-AF transition, but the $d_{3z^2-r^2}$ orbital population significantly
increases in the FM phase. This agrees with conclusions
from magnetic Compton
scattering measurements~\cite{koizumi01}. Further, our results show a
striking cooperative effect between Jahn-Teller distortion and magnetic
order. Indeed, the evolution of the occupations with $x$ in a
given magnetic phase reflects the change of the apical Mn-O
bond lengths~\cite{kubota00}.
But the crossover due to the change in magnetic order
(dotted line) is largely due to the change in bilayer splitting.

\begin{figure}
\includegraphics[width=\hsize]{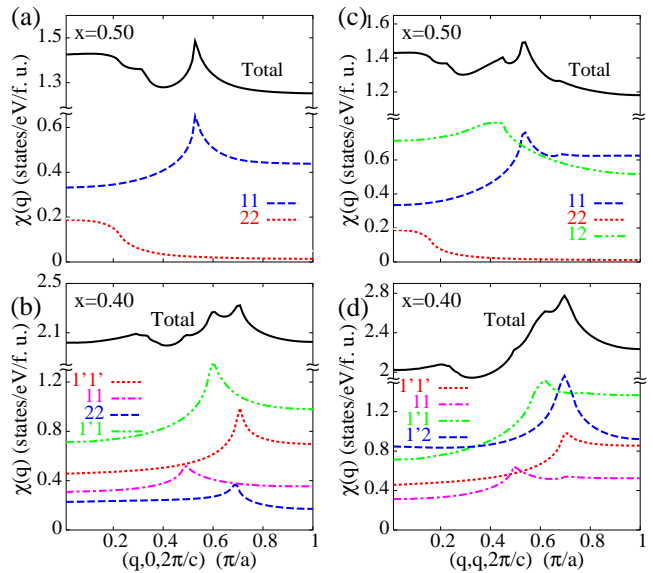}
\caption{\label{fig4} (color online)
Static susceptibility $\chi({\bf q})$ for {\bf q} along two symmetry
directions for $x=0.5$ (AF) and $x=0.4$ (FM). The labels indicate
which band transitions are responsible for the structure in $\chi$.
Labeling corresponds to that in
Fig.~\ref{fig2}. Note the break in the vertical axis.}
\end{figure}

The above results also indicate that the possible nesting instabilities
suggested to play an important role in these
materials ~\cite{chuang01,kim07} have a non-trivial dependence on $x$.
To see this, we calculated the generalized charge
susceptibility (constant matrix element approximation)
\begin{equation}
\chi({\bf q},\omega\rightarrow 0)={1\over N}\sum_{n,m}
\sum_{{\bf k},\sigma}
{{f_{n{\bf k}\sigma}(1-f_{m{\bf k}+{\bf q}\,\sigma})}
\over {\epsilon_{m{\bf k}+{\bf q}\,\sigma}-\epsilon_{n{\bf k}\sigma}-\omega}}.
\end{equation}
For this calculation, we
again used 32340 {\bf k}-points in the irreducible wedge and
the tetrahedron method with linear
interpolation~\cite{rath75}. In Fig.~\ref{fig3}, we show
$\chi({\bf q})$ for
{\bf q} of the form $(q_x,q_y,2\pi/c)$. This is of interest because
diffuse x-ray scattering data have shown peaks for $q_z=2\pi/c$ in the
$x=0.4$ compound~\cite{vasiliu99}. Both the AF and
FM cases show clear peaks and kinks 
correlating closely with the topology of the Fermi surface.
The calculations were done taking into account only the bands crossing 
$E_{\rm F}$. We verified that including more bands does not affect
significantly the momentum structure of the response. Also, the $e_g$ to
$t_{2g}$ transitions were
ignored to crudely simulate the neglected matrix elements effects.
We note that the $t_{2g}$ transitions mostly contribute to a
diffuse response centered at $\Gamma$, so
we focus on the $e_g$ transitions only.

Figure~\ref{fig4} takes a closer look for {\bf q} along two directions.
For ${\bf q}=(q,0,2\pi/c)$,
Fig.~\ref{fig4}(a) (AF) shows a 
strong peak at
$q=0.53$ (11 transition). The interband transition (12) has little structure
(not shown). Figure~\ref{fig4}(b)
(FM) shows two main peaks, for $q=0.62$ 
(1'1 transition), and 0.71 (1'1' and 22 transitions). The peak arising
from 11 transitions is relatively weaker in the FM case.
It is significant
that the 1'1 peak at this doping level matches
the bond centered peak reported at $q=0.6$ in
Ref.~\onlinecite{vasiliu99}.
For ${\bf q}=(q,q,2\pi/c)$,
Fig.~\ref{fig4}(c) (AF) shows the same (11) peak
as Fig.~\ref{fig4}(a), and a less well defined maximum
at $q=0.42$ (12 transition).
This maximum is equally ill-defined
at other doping levels.
Fig.~\ref{fig4}(d)
(FM) shows a clearly dominant peak at $q=0.7$ arising from
mainly 1'1' and 1'2 transitions. We point out that
the two kinks for smaller
$q$ (1'1 and 11) are weaker at lower doping levels
because of strong broadening of these features.

\begin{figure}
\includegraphics[width=\hsize]{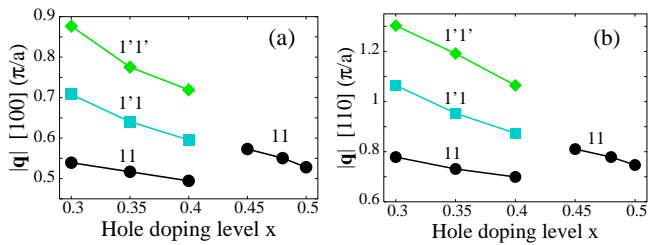}
\caption{\label{fig5} (color online)
Nesting vector lengths (extracted from the peaks in $\chi$) as a
function of doping level $x$ for {\bf q} along (a) the [100]
and (b) [110] directions.}
\end{figure}

Finally, in Fig.~\ref{fig5} we consider the
dependence on $x$ of the lengths of the magnitude of the nesting vectors
along the [100] and [110]
directions, inferred previously from photoemission~\cite{chuang01,sun06}.
We extracted these from the peaks in $\chi$
corresponding to those in Fig.~\ref{fig4}.
Again, we remark that
we do not have results for the actual canted phase for $0.45\alt x<0.5$.
Note that the 11 vector for $x=0.4$ in the [110] direction
matches that of the so-called CE-ordering wave
vector~\cite{chuang01}.  
We particularly point to the jump of the 11 nesting
vector when the magnetic phase changes. Clearly,
this is due to the `turning on' of the bilayer splitting in the FM phase.
Again,
this illustrates  dramatically the correlation between $d_{3z^2-r^2}$ and
$d_{x^2-y^2}$ orbital mixing, bilayer splitting, and FM order.
Further, this jump
may explain the non-monotonic $x$-dependence of the nesting
vector recently inferred from photoemission and diffuse scattering
measurements~\cite{stephan}.
The charge ordering peaks observed in the latter
are either along the bond direction,  the diagonal, or both,
depending on the doping level.  Typically, the diagonal response is largest,
since although the nesting is not as good for such vectors, twice as many
Fermi surface faces
are brought into coincidence as compared to the bond oriented
case~\cite{schulz}. We also note that in the FM phase, for transitions
involving 1',
the peak structure for $q_z=0$ along the bond direction
is less distinct than for $q_z=2\pi/c$.
This is because of the `tilting' of the 1' surface (a consequence
of its strong $d_{3z^2-r^2}$ admixture), which alternates in direction
between
$q_z=0$ and $2\pi/c$ due to the c-axis dispersion.  For the same reason, the
diagonal peaks involving 1' are stronger for $q_z=0$.
This may explain why the diffuse
scattering peaks occur for $q_z=2\pi/c$ for the bond directions and $q_z=0$
for the diagonal directions~\cite{vasiliu99,stephan}.

In summary, we find that mixing of the Mn $d_{x^2-y^2}$ and $d_{3z^2-r^2}$
orbitals and bilayer splitting
play a fundamental role in the electronic structure of the bilayer manganites,
particularly in the FM phase, and correlate closely with the transition from
FM to AF order with doping.
These two effects result in different Fermi surface topologies for the two
phases.
As a result, the static susceptibility, and the $e_g$ orbital polarization,
are predicted to
have a non-monotonic dependence on doping due to the change from FM to
AF order as the hole doping level increases.

We acknowledge the support of the US DOE, Office of Science, under Grant 
No.~DE-FG02-88ER45372 and Contract No.~DE-AC02-06CH11357,
and a computer time grant at the NERSC. We thank B. Barbiellini,
L.-H. Ye, J.-H. Song, J. F. Mitchell, R. Osborn, S. Rosenkranz, U. Chatterjee
and J. C. Campuzano for useful discussions.


\begin{thebibliography}{99}
\bibitem{comment0} In the FM case, the states of $d_{xy}$
symmetry (minority spin) result in two slightly split cylinders (3 and 3')
around $\Gamma$ [not shown in Fig.~\ref{fig2}(b)].
\bibitem{comment1} Bilayer splitting at $x$=0.4 is observable in magnetic
Compton profiles~\cite{mijnarends07}.  Some differences between bulk probes
and photoemission are expected given the latter's surface sensitivity.
\end{thebibliography}
\end{document}